\documentclass[preprint]{elsarticle}
\makeatletter
\def\ps@pprintTitle{%
\let\@oddhead\@empty
\let\@evenhead\@empty
\def\@oddfoot{}%
\let\@evenfoot\@oddfoot}
\makeatother

\usepackage{graphicx}
\usepackage{caption}
\usepackage{subcaption}
\graphicspath{{./figures/}{./}}

\usepackage{amssymb}
\usepackage{amsthm}
\usepackage{amsmath}
\usepackage[]{algorithm2e}

\usepackage{lineno}
\usepackage{hyperref}
\hypersetup{
    colorlinks=true,
    linkcolor=blue,
    filecolor=magenta,      
    urlcolor=cyan,
}
\modulolinenumbers[5]
\usepackage{layouts}

\newcommand*{\myeqref}[2][Eq.~]{%
  \hyperref[{#2}]{#1(\ref*{#2})}%
}

\begin{document}

\begin{frontmatter}



\title{The stochastic counterpart of conservation laws with heterogeneous
conductivity fields: application to deterministic problems and uncertainty quantification}

\author[farm]{Amir H. Delgoshaie \corref{cor}}
\ead{amirdel@stanford.edu}
\author[mse]{Peter W. Glynn}
\ead{glynn@stanford.edu}
\author[eth]{Patrick Jenny}
\ead{jenny@ifd.mavt.ethz.ch}
\author[farm]{Hamdi A.~Tchelepi}
\ead{tchelepi@stanford.edu}
\address[farm]{Department of Energy Resources
Engineering, Stanford University. 367 Panama Street, Stanford, CA, 94305, USA}
\address[mse]{Department of Management Science and Engineering,
Stanford University.
475 Via Ortega, Stanford, CA, 94305, USA}
\address[eth]{Institute of Fluid Dynamics, ETH Z\"urich.
Sonneggstrasse 3, CH-8092 Z\"urich, Switzerland}
\cortext[cor]{Corresponding author}

\begin{abstract}
Conservation laws in the form of elliptic and parabolic partial differential equations (PDEs) are
fundamental to the modeling of many problems such as heat transfer and flow in porous media.
Many of such PDEs are stochastic due to the presence of uncertainty in the conductivity field.
Based on the relation between stochastic diffusion processes and PDEs, Monte Carlo (MC) methods are
available to solve these PDEs.  These methods are especially relevant for cases where we are
interested in the solution in a small subset of the domain. The existing MC methods based on the
stochastic formulation require restrictively small time steps for high variance conductivity fields.
Moreover, in many applications the conductivity is piecewise constant and the existing methods are
not readily applicable in these cases. Here we provide an algorithm to solve one-dimensional
elliptic problems that bypasses these two limitations.  The methodology is demonstrated using
problems governed by deterministic and stochastic PDEs.  It is shown that the method provides an
efficient alternative to compute the statistical moments of the solution to a stochastic PDE at any
point in the domain. A variance reduction scheme is proposed for applying the method for efficient
mean calculations. 

\end{abstract} 
 \end{frontmatter}


\section{Introduction}
Elliptic and parabolic partial differential equations (PDEs) arise in many applications in
describing conservation laws. The mass conservation equation resulting from Fick's law, the heat
equation, and the pressure equation in the context of flow in porous media are some of the prominent
examples of such applications. These equations have the following general form 

\begin{equation}\label{parabolic}
\nabla . (K\nabla p) = c\frac{\partial p}{\partial t},  \\
\end{equation}
when the problem is time dependent and

\begin{equation}\label{elliptic}
\nabla . (K\nabla p) = 0,  
\end{equation}
when the problem is at steady state. Here, $p$ is the unknown (concentration, temperature, or pressure)
and $K$ is the conductivity tensor. In these problems, the flux (of mass or heat) is governed
by a Fick's type law, i.e,

\begin{equation}
 q = -K\nabla p.  
\end{equation}

In many practical settings, closed-form solutions of PDEs~\eqref{parabolic} and \eqref{elliptic} do not
exist, and numerical methods such as finite-volume (FV) and finite-element methods are used to
compute numerical solutions of these PDEs \cite{versteeg1995computational}. These numerical methods
rely on discretization of the PDE for the domain of interest and deriving a set of linear equations
for the solution of the PDE on the discretized grid. Once the system is reduced to a linear system,
efficient numerical methods can be employed to solve it \cite{trefethen1997numerical}.
These linear solvers compute the solution for all grid points simultaneously.\\

In stochastic modeling, elliptic and parabolic PDEs that are similar to equations~\eqref{parabolic}
and \eqref{elliptic} arise when calculating various expected values for a stochastic diffusion
process \cite{oksendal2003stochastic}.  In the stochastic modeling nomenclature, these PDEs are
referred to as backward equations, and the differential operator describing the left-hand-side of
equations~\eqref{parabolic} and \eqref{elliptic} is referred to as $\mathcal{L}$. In the view of the
connection between diffusion processes and PDEs (the Feynman-Kac formulation), the stochastic
counterpart of $\mathcal{L}$ has been used in methods such as backward walks and random walks on
boundary \cite{pardoux1998backward, sabelfeld1994random} to solve elliptic and
parabolic PDEs. Moreover, methods based on the Feynman-Kac formulation are available that can handle
Dirichlet, Neumann, and Robin boundary conditions \cite{buchmann2003solving, hu1993probabilistic,
zhou2016numerical}.\\

Recognizing that equations \eqref{parabolic} and \eqref{elliptic} correspond to backward equations of
specific stochastic processes has multiple advantages.  First, unlike solving linear systems, by
using the stochastic representation of the problem, the solution for any subset of points in the
domain can be found independently of the solution at points outside the subset of interest. Second,
the numerical solution can be computed at any point in the domain without the need for a mesh.
Third, the stochastic solution strategy is `embarrassingly parallel', which allows for efficient
implementations that can take full advantage of GPUs and massively parallel CPUs.\\

In many applications, such as flow in natural porous formations, the conductivity field,
$K$, is highly heterogeneous. This heterogeneity poses additional challenges for using the
stochastic counterpart of $\mathcal{L}$ in Monte Carlo algorithms. At the same time, there is often
uncertainty associated with $K$. The advantages of stochastic formulations become more
relevant in the case where  $K$ is modeled as a random field. The most common way to find
the solution under uncertainty in the conductivity field is to solve the equations numerically for
an ensemble of conductivity realizations to compute the statistical moments, or the distribution, of
the unknown variable. This can be a very expensive computational procedure. Using the stochastic
formulation, the moments of the solution or the one-point probability density function (PDF) at any
location in the domain is obtained by computing the solution only at the point of interest -
independently of other points - for the different realizations of the conductivity field.\\

Anker et.~al \citep{anker2017sde} recently used the Feynman-Kac formulation to solve elliptic
problems in heterogeneous conductivity fields. They also proposed a method to efficiently find the
mean solution in random heterogeneous domains. In the stochastic counterpart of $\mathcal{L}$ for
heterogeneous conductivity fields, the gradient of conductivity appears as the drift function, so
the method is valid only for smooth conductivity fields. This is a good assumption in many scenarios
where $K$ is indeed sufficiently smooth or it can be well approximated by smooth functions.
This is true both for deterministic and stochastic $K$, where the conductivity can be
represented by truncated Fourier series or Karhunen-Loeve (KL)
expansions~\cite{woods2000probability}.\\

There are additional challenges remaining for using the stochastic counterpart of
equations~\eqref{parabolic} and~\eqref{elliptic}. In many practical applications the conductivity field 
is piecewise
constant and the method proposed in \cite{anker2017sde} would not be applicable in those
settings. Moreover, even when solving elliptic problems with the stochastic formulation, the
location of the computational particles are incremented in pseudo time steps. For highly
heterogeneous conductivity fields which are common in porous media applications, the method would be
inaccurate unless for very small time steps.  In general, the correlation length and the variance of
the conductivity field determine the right time step size. For some cases this would
render the stochastic approach computationally too expensive.\\

In this work, we first review the correspondence between $\mathcal{L}$, the operator in the backward
equations of diffusion processes, and the differential operator in the steady-state and transient
conservation equations with heterogeneous coefficients. Next, a stochastic algorithm is
proposed to solve the elliptic problem in one dimension that would be valid for
piecewise constant conductivity fields. We provide a heuristic proof of the method based on the
exit probability of diffusion processes. A rigorous proof of the proposed method can be provided
based on skew Brownian motion, similar to~\cite{harrison1981skew, lejay2006scheme}. In
one dimension, the proposed method is exact and is not sensitive to the variance or the correlation
length of the conductivity field.\\

The paper is organized as follows. In section~\ref{review} the Feynman-Kac formulation for
heterogeneous media is reviewed. Potential issues for using the stochastic formulation in high-variance conductivity fields is discussed in section~\ref{sec-problems}. The stochastic algorithm
for solving elliptic PDEs with piecewise constant coefficients is presented in
section~\ref{peicewise} and a deterministic example is provided in section~\ref{sec-example}.
Various examples are provided for using the proposed stochastic algorithm for uncertainty
quantification in section~\ref{sec-uq}. A variance reduction scheme for calculating the mean
solution to a stochastic PDE is discussed. Finally, conclusions and possible extensions 
of the algorithm are discussed in section~\ref{sec-conclusions}.\\ 


\section{Conservation laws versus backward equations}
\label{review}
Consider $X(t)$ that satisfies the following stochastic differential equation (SDE)
\begin{equation}\label{SDE}
dX(t) = \mu(X(t))dt + 
\sigma(X(t))dB(t). 
\end{equation}
We are interested in computing various expected values of functions of $X(t)$ conditional on
starting the process at $X(0)=x$. It is standard to adapt the following notation
\begin{equation}
E_x[\cdot] = E[\cdot | X(0) = x]. 
\end{equation}
As it is shown in~\cite{oksendal2003stochastic, gardiner2009stochastic},
\begin{equation}
u^*(t,x) =E[r(X(t))|X(0) = x] =
E_{x}[r(X(t))], 
\end{equation}
which is the expected value of $r(X(t))$ if we start from $X(0)=x$, can be found by solving the following parabolic PDE
\begin{equation}\label{backward}
\begin{gathered}
\mathcal{L}u^*(x,t) = \frac{\partial u^*(x,t)}{\partial t} \\
u^*(x, 0) = r(x).
\end{gathered}
\end{equation}
In equation~\eqref{backward} the differential operator $\mathcal{L}$ is defined as follows
\begin{equation}
\mathcal{L} = \sum_{i=1}^{d}\mu_i(x)\frac{\partial}{\partial x_i} + 
\frac{1}{2}\sum_{i,j=1}^{d}b_{ij}(x)\frac{\partial^2 }{\partial x_i\partial x_j}.
\end{equation}
Here $b_{ij}(x)$ refers to the elements of $\sigma(x)\sigma(x)^T$.
In these equations, $X(t), \mu(X(t)) \in \mathbb{R}^d$ and $\sigma(X(t)) \in \mathbb{R}^{d
\times d}$.  $B(t) \in \mathbb{R}^d$ is a $d$-dimensional Brownian motion. 
To  be concrete, in two dimensions
\begin{equation}
\begin{gathered}
\begin{bmatrix}
b_{11}(x) & b_{12}(x)\\ 
b_{21}(x) & b_{22}(x)\\ 
\end{bmatrix} = 
\sigma(x) \sigma^T(x)= \\
\begin{bmatrix}
\sigma_{11}^2(x) + \sigma_{12}^2(x) & \sigma_{11}(x)\sigma_{21}(x) +\sigma_{12}(x)\sigma_{22}(x)\\ 
\sigma_{11}(x)\sigma_{21}(x) +\sigma_{12}(x)\sigma_{22}(x) & \sigma_{22}^2(x) + \sigma_{21}^2(x) \\ 
\end{bmatrix}.  
\end{gathered}
\end{equation}
As shown in~\cite{gardiner2009stochastic}, the rigorous connection between SDEs and PDEs involves
the use of Ito's formula. In order to apply Ito's formula to show the connection between SDEs and
PDEs, the solution to the PDE needs to be continuously differentiable in time,
and piecewise twice continuously differentiable in the spatial variable. In this section we assume
that the solutions to the considered PDEs satisfy the necessary conditions for using applying Ito's
formula.\\

The connection between equations~\eqref{SDE} and \eqref{backward} makes it possible to solve a PDE
similar to equation~\eqref{backward} using MC simulation of its SDE counterpart.  The
algorithm is as follows: to find $u^*(x,t)$, launch random walks from $x$ at time
$t$; evolve them according to equation~\ref{SDE} for $t$ time units backward in time (to $t=0$),
and store $r(X(0))$. The solution is then the average of the stored values $r(X(0))$.  For
homogeneous isotropic conductivity fields ($K_{ij}(x) = K\delta_{ij}$), this method is referred to as 
`backward walks' \cite{chorin2009stochastic}.\\

Similarly, the expected value
\begin{equation} \label{exp-hitting}
u^*(x) = E_x[r(X(T))], 
\end{equation}
where $T = \inf\{t\geq0:X(t)\in C^c\}$ is the first hitting time of $C^c$ (complement of set $C$), 
can be found by solving the following elliptic PDE

\begin{equation}\label{walkonb}
\begin{gathered}
\mathcal{L}u^*(x) = 0, ~ x \in C\\
u^*(x) = r(x) ,~ x \in C^c.
\end{gathered}
\end{equation}
Hence, to solve a PDE similar to equation~\eqref{walkonb}, one can use MC simulations of
equation~\eqref{SDE}.  The algorithm is as follows:  to find $u^*(x)$, start many
random walks from $x$, and evolve them according to equation~\eqref{SDE} until they hit the boundary
for the first time, then store the boundary value at the hitting times. 
The solution can be obtained by averaging these boundary values.  
For cases where the conductivity field is homogeneous, this
method has been developed and is referred to as random walk on boundary \cite{sabelfeld1994random}.\\

In order to use ``backward walks'' and ``random walks on boundary'' for problems where the
conductivity field is heterogeneous, we need to find the stochastic counterpart of the differential
operator in equations~\eqref{parabolic} and \eqref{elliptic}.  In the following section, we
illustrate this stochastic representation by expanding the conservation equations and comparing the
expanded operator with $\mathcal{L}$. The comparison is performed for a two-dimensional system;
however, the argument can readily be extended to n dimensions.

\subsection{Comparison of the differential operators}

Since the linear operator for both elliptic and parabolic problems is the same, we focus on the
elliptic problem. Assuming $K$ is differentiable, expansion of equation~\eqref{elliptic}
leads to

\begin{equation}\label{expanded2}
\begin{gathered}
\frac{\partial K_{11}}{\partial x_1}\frac{\partial p}{\partial x_1} +  
\frac{\partial K_{22}}{\partial x_2}\frac{\partial p}{\partial x_2} + 
\frac{\partial K_{12}}{\partial x_1}\frac{\partial p}{\partial x_2} +  
\frac{\partial K_{21}}{\partial x_2}\frac{\partial p}{\partial x_1} + \\
K_{11}\frac{\partial^2 p}{\partial x_1^2}+ K_{22}\frac{\partial^2 p}{\partial x_2^2} + 
K_{12}\frac{\partial^2 p}{\partial x_1 \partial x_2}+ 
K_{21}\frac{\partial^2 p}{\partial x_2 \partial x_1} = 0.
\end{gathered}
\end{equation}
Equation~\eqref{expanded2} can be written as $\mathcal{L}_1p = 0$, where

\begin{equation}
\begin{gathered}
\mathcal{L}_1 = 
\Big(
\Big(\frac{\partial K_{11}}{\partial x_1} + \frac{\partial K_{21}}{\partial x_2}\Big)
\frac{\partial }{\partial x_1} +  
\Big(\frac{\partial K_{22}}{\partial x_2} + \frac{\partial K_{12}}{\partial x_1}\Big)
\frac{\partial }{\partial x_2}
\Big) + \\
\Big(
K_{11}\frac{\partial^2 }{\partial x_1^2}+ K_{22}\frac{\partial^2 }{\partial x_2^2} + 
K_{12}\frac{\partial^2 }{\partial x_1 \partial x_2} + 
K_{21}\frac{\partial^2 }{\partial x_2 \partial x_1}
\Big).
\end{gathered}
\end{equation}
The first part of $\mathcal{L}_1$ matches the drift term of the backward operator. Note that the derivatives
of the conductivity field constitute the drift term, or the preferential direction for the random
walks. In two dimensions, comparison of $b_{ij}$ and the second-order term in $\mathcal{L}_1$ yields

\begin{equation}\label{offdiag}
\begin{bmatrix}
K_{11} & K_{12} \\ 
K_{21} & K_{22} \\
\end{bmatrix} = \frac{1}{2}
\begin{bmatrix}
\sigma_{11}^2 + \sigma_{12}^2 & \sigma_{11}\sigma_{21} +\sigma_{12}\sigma_{22}\\ 
\sigma_{11}\sigma_{21} +\sigma_{12}\sigma_{22} & \sigma_{22}^2 + \sigma_{21}^2 \\ 
\end{bmatrix}  .
\end{equation}
System~\eqref{offdiag} consists of four equations and four unknowns. For symmetric $K$, this system
can easily be solved to find $\sigma_{ij}$ corresponding to $\mathcal{L}_1$.

\section{Challenges in using the stochastic formulation for high
variance conductivity fields}
\label{sec-problems}

\begin{figure}
\begin{center}
\includegraphics[width=0.6\linewidth]{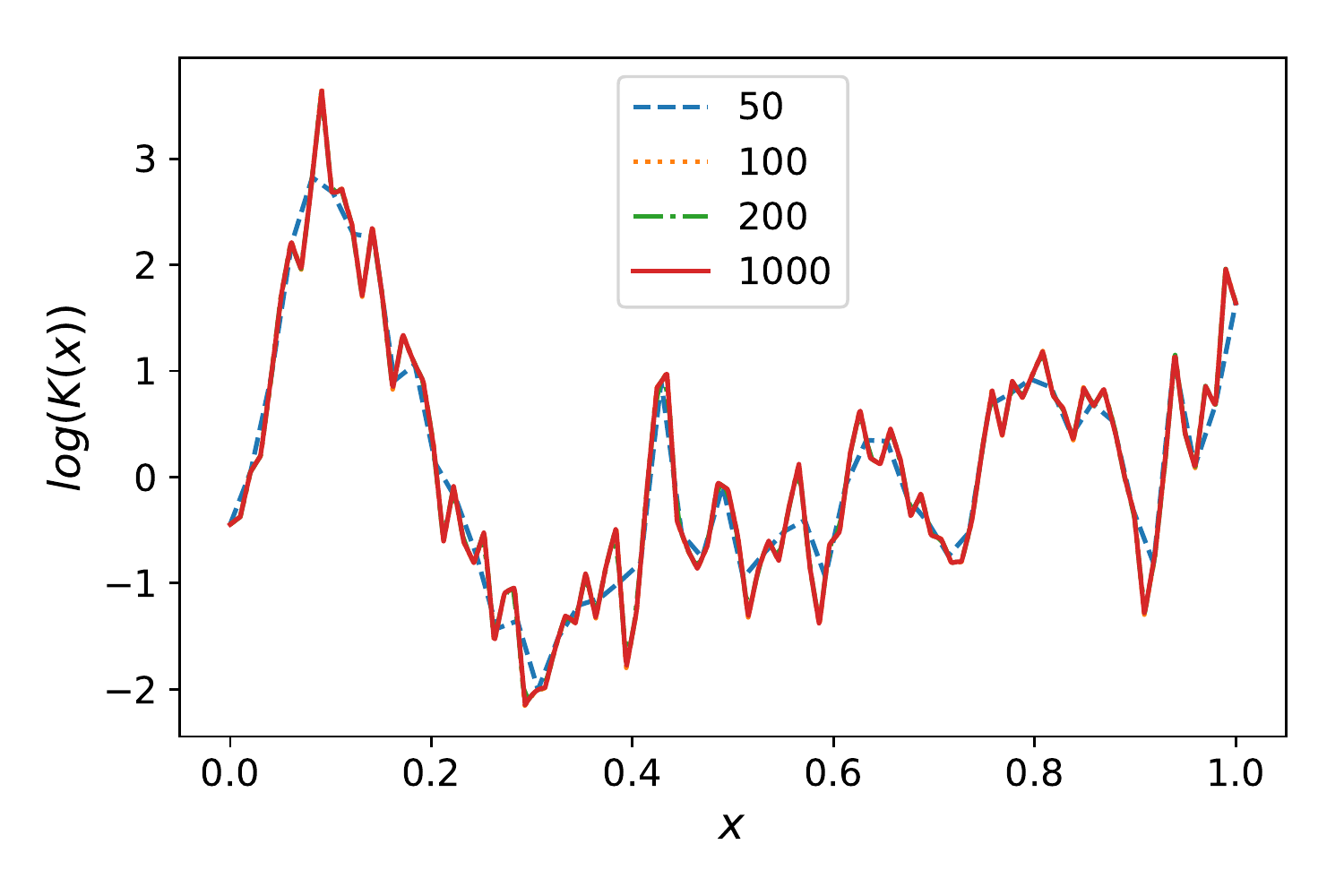}
\caption{One realization of $\log(K(x))$ described in section~\ref{sec-problems}, 
evaluated with four different resolutions. 
\label{realization}} 
\end{center} 
\end{figure}

Here, we consider a one-dimensional case, where $K(x)$ is one realization of a conductivity field
with a log-normal distribution with mean zero and a variance of four ($\log(K) \sim
\mathcal{N}(0,4)$), with an exponential covariance function. Such conductivity fields are very
common in porous media applications.  The domain is $D = [0,1]$ and the boundary conditions are
$p(0)=1$ and $p(1)=0$.  The dimensionless correlation length ($l_Y$) of $K(x)$ is equal to $0.05$.
We draw this realization using the Karhunen-Loeve (KL) expansion of the conductivity field. By
truncating the KL expansion after $91$ terms, to capture $99\%$ of the energy of the field, 
we ensure that the conductivity is differentiable and the formulation
discussed in the previous section is applicable.  
Figure \ref{realization} shows one realization of $\log(K)$ generated from this KL expansion with
multiple different resolutions.
The code available at \cite{uq-course} was used
for generating the KL expansion.\\  



\begin{figure}
\begin{center}
\includegraphics[width=\linewidth]{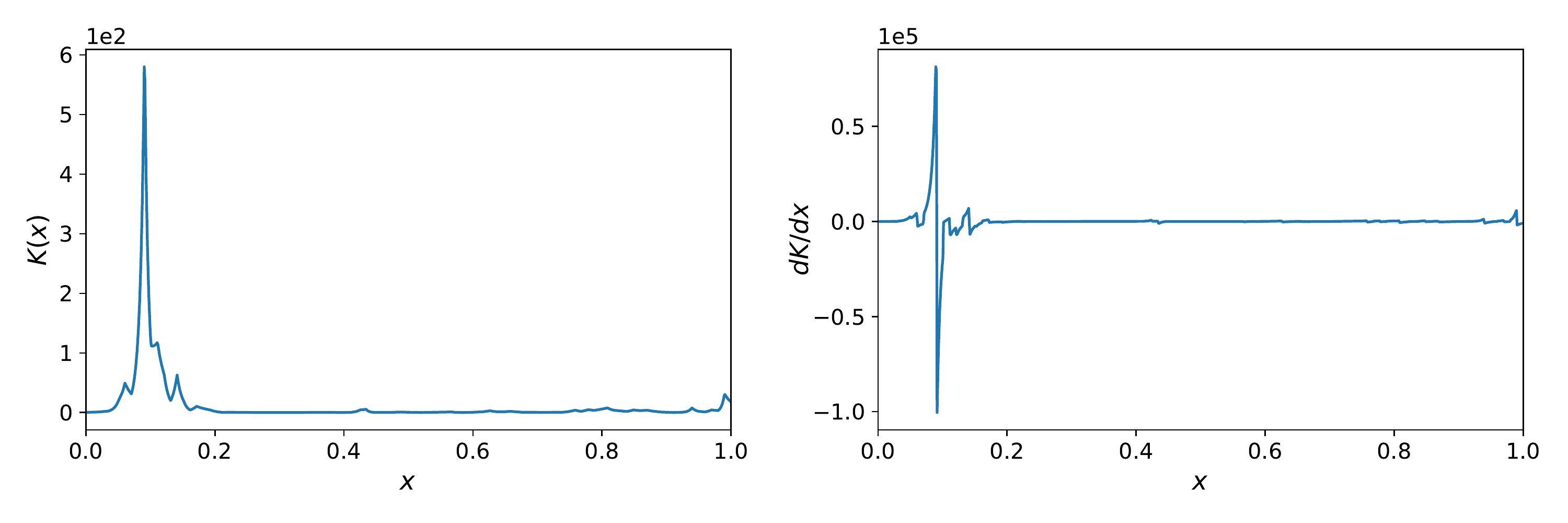}
\caption{One realization of the permeability field described in section~\ref{sec-problems} along with its derivative.
The derivative is used as the drift term in the stochastic counterpart of the elliptic problem. 
\label{perm}} 
\end{center} 
\end{figure}

\begin{figure}
\begin{center}
\includegraphics[width=0.6\linewidth]{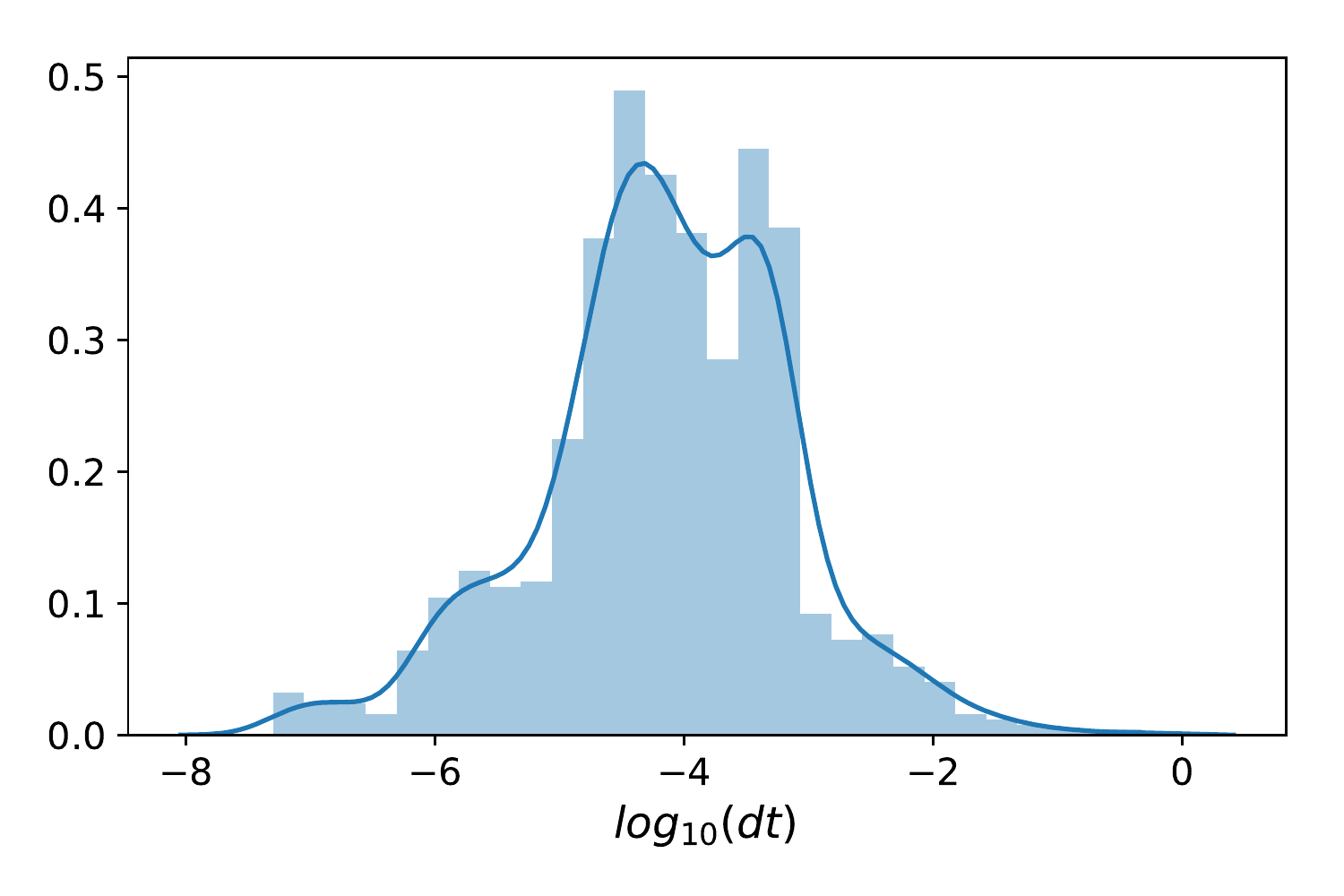}
\caption{Distribution of $dt$ that would satisfy $\frac{dK}{dx}dt< 0.1l_Y$.
\label{cfl}} 
\end{center} 
\end{figure}

Figure \ref{perm} shows the realization of the permeability field used in this
example and its derivative.  To ensure that we have resolved the permeability
field sufficiently, $K(x)$ was evaluated at $n=1000$ equidistant points. This
example illustrates a potential challenge for solving the elliptic problem with
the Feynman-Kac formulation even when the conductivity is differentiable. 
The derivative of $K(x)$ is a drift velocity.  Starting at a point
with a very high value of the drift function, the time step should be selected
such that the particle can still see the variations in $K$. More specifically, one should resolve
the smallest wavelength in the truncated KL expansion, which typically requires resolving a length scale 
$\ll l_Y$.  
The distribution of the time step
that would satisfy $\frac{dK}{dx}dt< 0.1l_Y$ is shown in Fig.~\ref{cfl}. This figure illustrates
that adaptive time stepping or a very small constant time step is required for ensuring that a
particle sees the variations in $K$ as it travels through the domain.  The time step restrictions
would become more strict with higher variance and smaller correlation length of $\log(K)$. Choosing
the right time step would also require the characterization on $dK/dx$. \\

Moreover, in many practical applications a piecewise constant conductivity is available on a grid.
In these scenarios $K$ is no longer differentiable and the methods proposed
in~\cite{anker2017sde} are not readily applicable.  In the next section we propose an algorithm for
solving the one-dimensional elliptic problem that bypasses the restrictions on $dt$ and extends
the application of the stochastic formulation to problems with piecewise constant $K$. 


\section{An algorithm for solving elliptic problems with piecewise constant $K$ in one dimension}
\label{peicewise}

Consider a 1D diffusion of the form 
\begin{equation}\label{SDE-1d}
dX(t) = \mu(X(t))dt + \sigma(X(t))dB(t). 
\end{equation}
As shown in~\cite{gardiner2009stochastic}, starting at point $x$, the probability of exiting an interval through its
right boundary before the left boundary is
\begin{equation}
P(T_r < T_l) = \frac{u(x)-u(l)}{u(r)-u(l)},
\end{equation}
where
\begin{equation}\label{eq-u}
u(x) = \int_{0}^{x} exp\Big(-\int_{0}^{y}\frac{2\mu(z)}{\sigma^2(z)}dz \Big)dy. 
\end{equation}
From section~\ref{review}, when $K$ is differentiable, we know that $\mu(z) =
dK(z)/dz$ and $\sigma(z) = \sqrt{2K(z)}$. Substituting into equation \eqref{eq-u} we obtain

\begin{equation}\label{eq-integral}
\begin{gathered}
u(x) = \int_{0}^{x} exp\Big(-\int_{0}^{y}\{\frac{d}{dz} \log(k(z))\} dz \Big)dy = \\
 \int_{0}^{x} exp\Big( -\log(K(y)) + \log(K(0))  \Big)dy = \\ 
K(0)\int_{0}^{x} exp\Big(\log(1/K(y))\Big) dy = 
K(0)\int_{0}^{x} \frac{1}{K(y)}dy.
\end{gathered}
\end{equation}
This shows that knowing $K$ is sufficient for calculating the exit
probability and $dK/dx$ does not appear in the final result.\\ 

Using equation~\eqref{eq-integral} we can find the exit probability for a two cell
problem, where the permeability of the left cell and the right cell are given
by $K_l$ and $K_r$. 
We can repeat the calculations in~\eqref{eq-integral} with a smooth approximation of the
conductivity, such that the drift function is defined everywhere in the two cell
problem.  Starting from the interface of the two cells, the probability of exiting from the right
boundary is equal to 

\begin{equation}\label{exitprob}
P(T_r < T_l) = \frac{1/K_l}{1/K_l + 1/K_r}.
\end{equation}
For a two cell problem with piecewise constant conductivity the presented proof is not rigorous.
This is the case since the conditions necessary for applying Ito's formula and establishing the
relation between the SDE and the PDE are not satisfied (see section~\ref{review}).  A rigorous proof
in that case can be provided by using skew Brownian motion and following an argument similar 
to~\cite{harrison1981skew,lejay2006scheme}.\\

From the stochastic formulation of the flow problem we know the pressure
solution at the interface, $p(x_I)$, is the expected value of the boundary
condition at the hitting time of the boundary  (see equation~\ref{exp-hitting}). For a two cell problem with boundary conditions 
$p_l$ and $p_r$, the interface pressure would be
\begin{equation}
\begin{gathered}
p(x_I) = E_{x_I}\big[p_T  \big] = \\
p_r  \frac{1/K_l}{1/K_l + 1/K_r} + p_l  \frac{1/K_r}{1/K_l + 1/K_r} = 
\frac{K_l p_l + K_r p_r}{K_l + K_r}. 
\end{gathered}
\end{equation}
Using a finite-volume method we arrive at the same solution for $p(x_I)$. In short, one can
calculate the flux going through the interface by using the conductivity of the
left cell and the right cell.
The flux going through the interface is equal to 
\begin{equation}
q = K_l \frac{p_l - p(x_I)}{\Delta x} = K_r \frac{p(x_I) - p_r}{\Delta x}
\end{equation}
from which we can find the interface pressure
\begin{equation}
p(x_I) = \frac{K_l p_l + K_r p_r}{K_l + K_r}.
\end{equation}

Following these observations we propose algorithm \ref{alg-piecewise} for
finding the solution to a one-dimesional elliptic problem with piecewise
constant coefficients. Since the method is exact, the only source of error in
algorithm~\ref{alg-piecewise}
is the Monte Carlo error related to the number of trajectories starting from the point where we seek
the solution.  

\begin{center}
\begin{algorithm}
\DontPrintSemicolon
$s = 0$\;
\For{$i\leftarrow 1$ \KwTo $M$}{
$x = x_0$\;
 \tcp{if $x$ is not a discontinuity point}
 \If{$x \not\in X_{discontinuity}$}{
      \tcp{Move the particle to one of the two closest discontinuity points with the 
      corresponding probability. $x_{l}$ and $x_{r}$ are the closest discontinuity points to the left and right of $x$.}
      $P_{l} = (x_{r}-x)/(x_{r}-x_{l})$\;
      \tcp{draw a uniform random variable $\alpha \in (0,1)$}
      \lIf {$\alpha < P_{l}$}{$x=x_{l}$}\lElse {$x=x_{r}$}
  }
 $Exit = $ checkExit($x$)\;
 \While {not $Exit$}{
     \tcp{advance the particle to the next discontinuity point}
	 $a_l = (x - x_{l})/K_l$\;
	 $a_r = (x_{r} - x)/K_r$\;
     $P_{r} = a_l / (a_r + a_l)$\;
     \tcp{draw a uniform random variable $\alpha \in (0,1)$}
     \lIf {$\alpha < P_{r}$}{$x=x_{r}$}\lElse{$x=x_{l}$}
     
     $Exit = $ checkExit($x$)\;
    }  
\tcp{add the value of the boundary condition to $s$}
$s = s+BC(x)$
} 
\Return $s/M$ \;
\BlankLine
 \caption{Algorithm for solving the elliptic problem with piecewise constant $K$.}
\label{alg-piecewise}
\end{algorithm}
\end{center}


\section{An illustrative deterministic example}
\label{sec-example}
\begin{figure}
\begin{center}
\includegraphics[width=\linewidth]{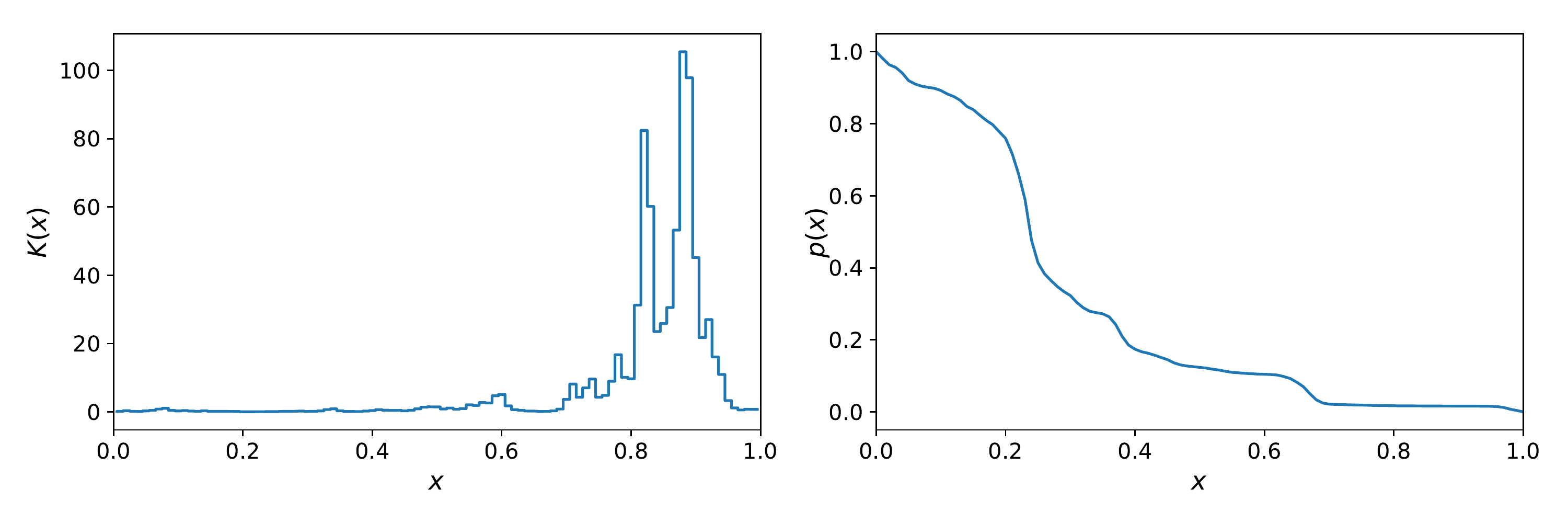}
\caption{Conductivity field and the corresponding pressure solution for the example in section~\ref{sec-example}
\label{e1-perm}} 
\end{center} 
\end{figure}

In this section, an example of using the proposed stochastic method for solving deterministic
problems is provided. We consider one realization of a Gaussian conductivity field where $\log(K)
\sim \mathcal{N}(0,4)$ with an exponential correlation function with $l_Y = 0.25$. Unlike
section~\ref{sec-problems}, the conductivity realization used here is a piecewise
constant field generated by the Fourier integral method~\cite{Pardo-Iguzquiza1993}.  The reference
solution, $p^*(x)$, for a one-dimensional elliptic problem can be found
analytically by integrating the one-dimensional version of equation~\eqref{elliptic}:
\begin{equation}\label{eq-analytic}
\begin{gathered}
\frac{d}{dx}\big(K(x) \frac{dp}{dx}\big) = 0, \;x\in (0,1)\\
s.t. \; p(0) = 1,\; p(1) = 0\\
p^*(x) = 1-\Big(\int_{0}^{1} \frac{dx}{K(x)}\Big)^{-1} \int_{0}^{x}\frac{dx}{K(x)}.
\end{gathered}
\end{equation}
The conductivity realization and the corresponding pressure solution are
illustrated in Fig.~\ref{e1-perm}.

\begin{figure}
\begin{center}
\includegraphics[width=0.6\linewidth]{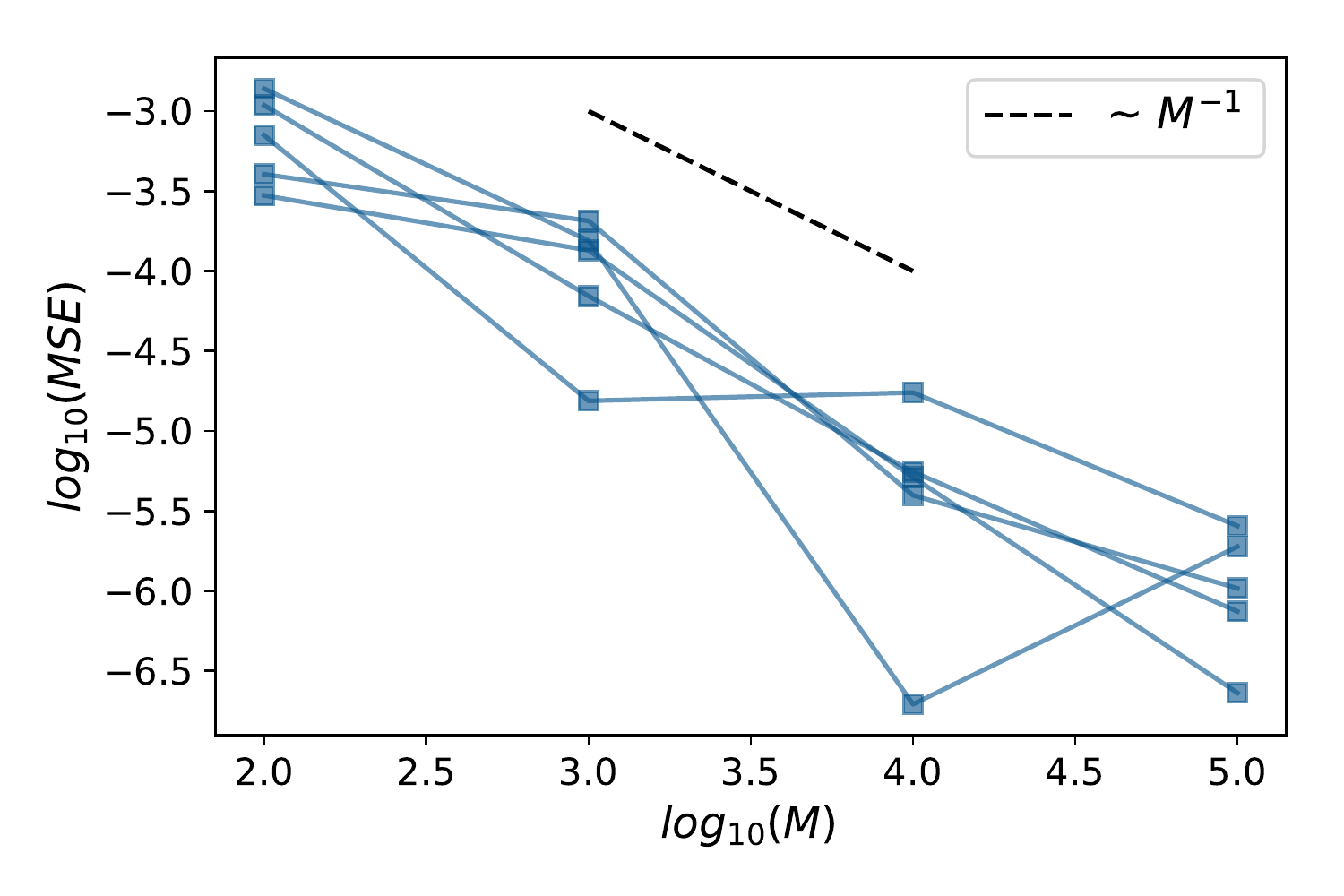}
\caption{Convergence of the solution obtained by algorithm~\ref{alg-piecewise}
to the analytical solution for the example in section~\ref{sec-example}. 
Different lines correspond to different experiments.  \label{e1-convergence}} 
\end{center} 
\end{figure}

The convergence of the MC solution obtained by using
algorithm~\ref{alg-piecewise} to the analytic solution is shown in
Fig.~\ref{e1-convergence} for different numbers of trajectories released per
point. The MC solution is calculated at all points $x \in S=\{0.25, 0.5, 0.75\}$. The mean square
difference is defined as
\begin{equation} 
MSE = \frac{1}{|S|}\sum_{x\in S} (p(x)-p^*(x))^2.
\end{equation}
Since the method is exact the only
source of error is the stochastic error due to the number of trajectories
followed per solution point. This is consistent with the $M^{-1}$ scaling of
the error, where $M$ is the number of random walks per point in $S$.

%
%

\section{Illustrative examples for uncertainty quantification}
\label{sec-uq}
Building on algorithm~\ref{alg-piecewise}, in this section we use ``backward walks on boundary'' to
quantify the uncertainty in the solution of elliptic PDEs with a random heterogeneous piecewise
constant conductivity. Algorithm~\ref{alg-uq} shows the modifications for uncertainty 
quantification. \\

\begin{figure}
\begin{center}
\includegraphics[width=\linewidth]{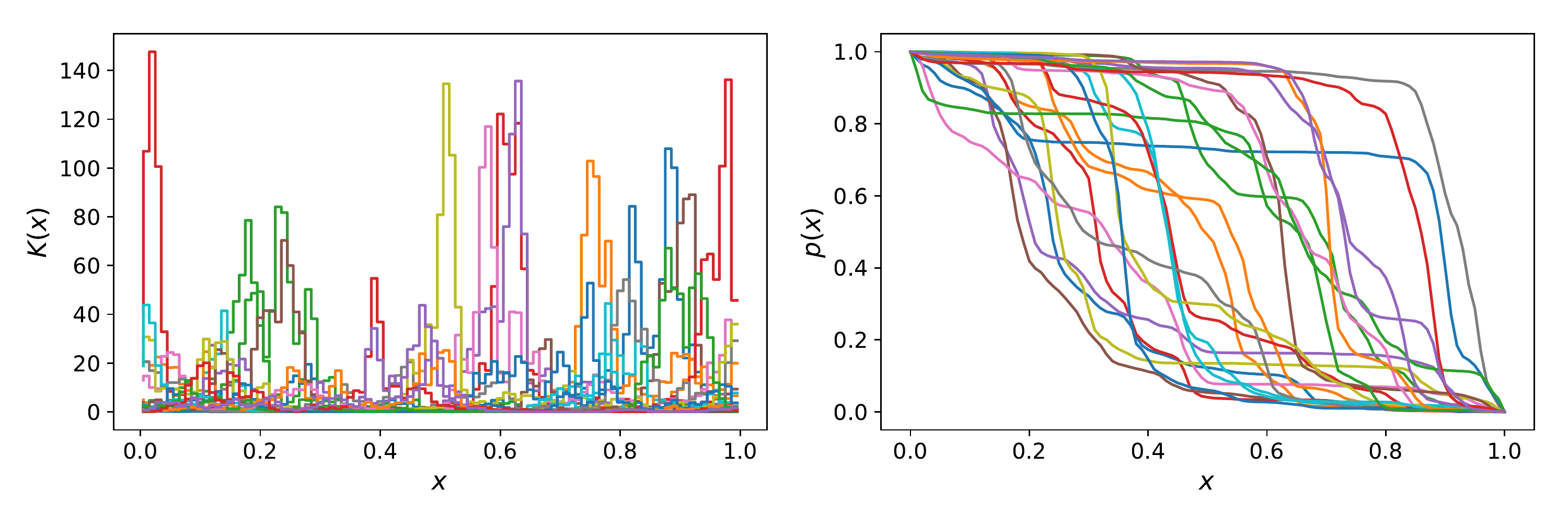}
\caption{Conductivity realizations sampled from the distribution described in section~\ref{sec-uq}
 and their corresponding analytical pressure solution.
\label{perm-uq}} 
\end{center} 
\end{figure}

In the examples provided in this section the conductivity field has an exponential correlation
structure with $l_Y=0.25$. The domain is $D = [0,1]$ and the boundary conditions are $p(0)=1$ and
$p(1)=0$.  Different realizations of the described conductivity field were generated and the flow
equation was solved using analytic integration (equation~\eqref{eq-analytic}) for all realizations.
Figure~\ref{perm-uq} illustrates a number of these realizations and their corresponding pressure
solution. The analytical solution for an ensemble of 100,000 realizations is used as the reference
solution in the following examples. 

\begin{algorithm}
\DontPrintSemicolon
\tcp{Initialization}
$S = [\; ]$\;
\tcp{Looping over different realizations of $K$}
\For{$i\leftarrow 1$ \KwTo $N$}{
    $K = $ generateField()\;
    \tcp{Sample from solution distribution using algorithm 1}
    $p = $ solveMC($x_0$, $K$, $M$)\;  
    $S$.store(p)
}
\tcp{return desired statistics (e.g. mean, histogram) using the estimated distribution 
of the solution}
\Return desiredStatistic($S$)\;
\BlankLine
 \caption{Uncertainty quantification using algorithm~\ref{alg-piecewise}.}
 \label{alg-uq}
\end{algorithm}

\subsection{Estimating the one-point distribution}
\label{one-point}
\begin{figure}
\begin{center}
\includegraphics[width=\linewidth]{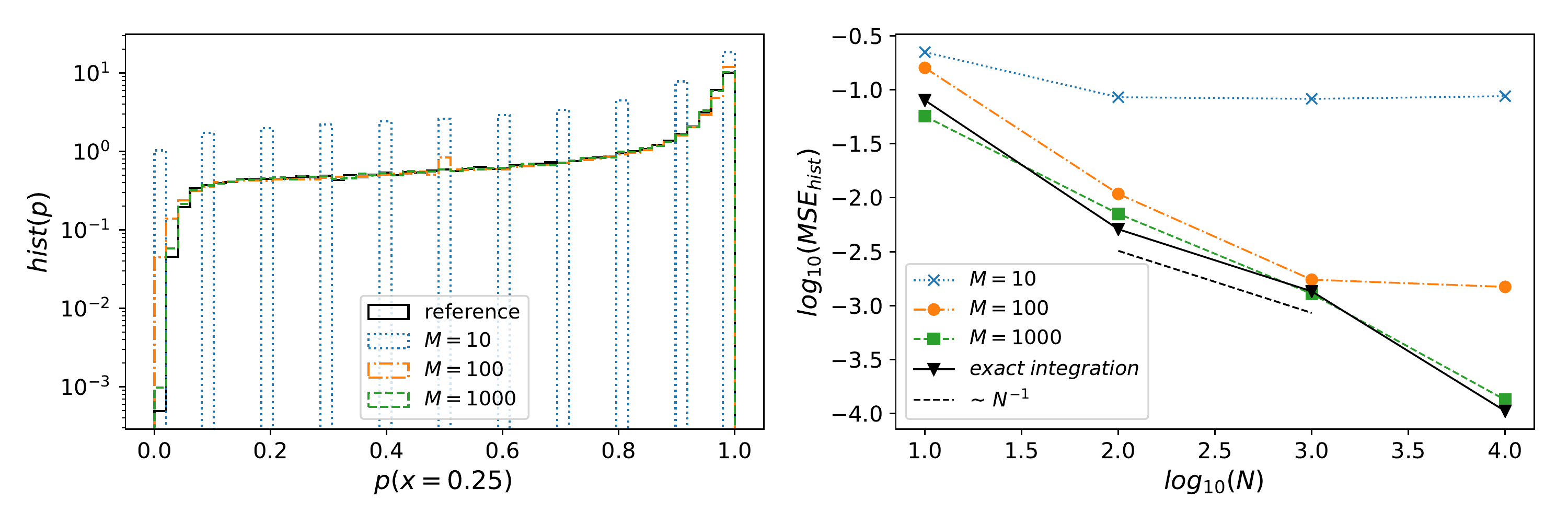}
\caption{Left: comparison of the one point histogram of pressure generated by
algorithm~\ref{alg-uq} and the reference histogram at $x=0.25$. Right:
Convergence of the histograms obtained from algorithm~\ref{alg-uq} to the
reference histogram.
\label{hist-uq}} 
\end{center} 
\end{figure}

In Fig.~\ref{hist-uq} the one-point histograms of $p(x)$ at $x=0.25$ generated by
algorithm~\ref{alg-uq} are compared with the reference histogram.  For generating the histograms in
Fig.~\ref{hist-uq}, $N=100,000$ realizations were used in algorithm~\ref{alg-uq}, and $M$ (the
number of random walks followed in each realization) was varied between $10$ and $1000$. The
histogram obtained from the analytical solution of $100,000$ realizations was used as the reference.
It can be observed that even for $M=100$ random walks per realization the obtained histogram is very
close to the reference.  Fifty equal width bins were used for all histograms in Fig.~\ref{hist-uq} .
We define mean square error for a histogram as
\begin{equation} 
MSE_{hist} = \frac{1}{n_{bins}}\sum_{i = 1}^{n_{bins}} (p(x_i)-p^*(x_i))^2.
\end{equation}
The convergence of the mean square error of the histograms obtained with algorithm~\ref{alg-uq}  to
the reference histogram is
compared with the convergence of the histogram calculated by analytic integration for different
number of realizations ($N$) in the right portion of Fig.~\ref{hist-uq}.



\subsection{Estimating the mean solution}
\label{mean-calc}
\begin{figure}[!h]
\begin{center}
\includegraphics[width=0.6\linewidth]{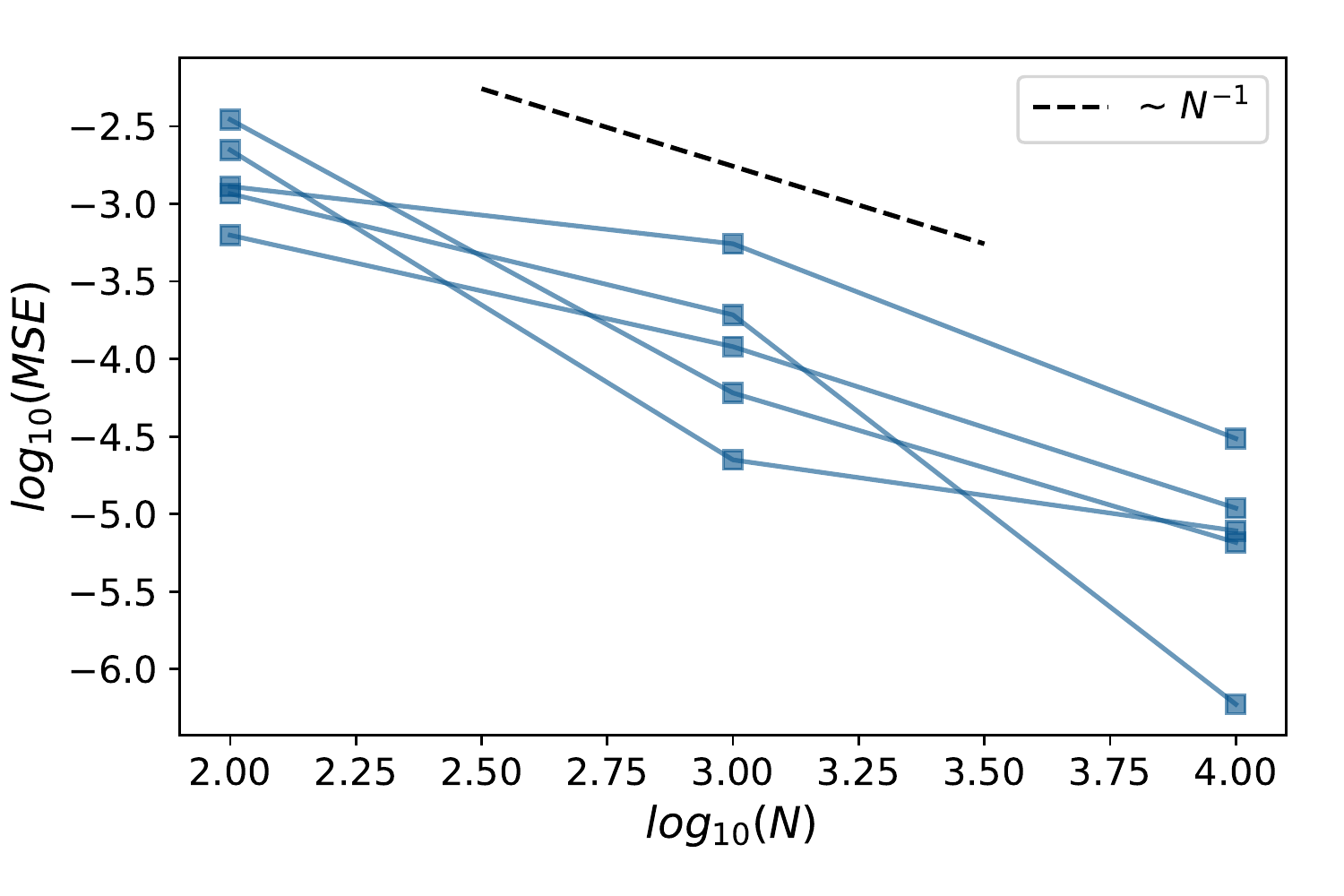}
\caption{Convergence of the mean solution calculated with algorithm~\ref{alg-uq} with $M=1$ to the
reference solution for the example described in
section~\ref{mean-calc}. Different lines correspond to different experiments. 
\label{mean-convergence}} 
\end{center} 
\end{figure}

Since the proposed method can be used for estimating the one-point distribution of the
solution, it can also be used for estimating the moments of the solution at any given point.
Estimating the mean is specifically efficient using algorithm~\ref{alg-uq}. As it was shown in
\cite{anker2017sde}, by tracking one trajectory per realization ($M=1$), the mean solution can be
calculated very efficiently .  This is the case, since the stochastic MC method provides an unbiased
estimate of the solution in each realization, and calculating the mean involves averaging the
solution of different realizations.  Figure~\ref{mean-convergence} illustrates the convergence of
the mean solution calculated with algorithm~\ref{alg-uq} at all points 
$x \in S=\{0.25, 0.5, 0.75\}$ to
the reference solution.  Here the mean square error is define as
\begin{equation} 
MSE = \frac{1}{|S|}\sum_{x\in S} \Big(\overline{p}(x)-\overline{p}^*(x)\Big)^2,
\end{equation}
where $\overline{p}(x)$ and $\overline{p}^*(x)$ are the mean solutions
calculated respectively using algorithm~\ref{alg-uq} and analytic
integration. These results illustrate that the proposed method can be used to
efficiently find the mean solution in highly heterogeneous conductivity fields. 



\subsection{Variance reduction for mean calculation}
\label{sec-trend}
In applications such as flow in porous media, it is common to have a trend in the log-conductivity
field. Based on the work in~\cite{gorji2015variance}, the algorithm proposed for mean calculation
can be modified to use the trend in the conductivity field to reduce the variance of the estimated
mean.
In short, for every realization of $K$, one could track a particle in that realization along with a 
shadow particle in the trend conductivity field using the same random numbers and store the boundary
conditions at the hitting points of the boundary for both particles. The contribution of that
realization to the mean would then be the sum of the difference between the boundary
conditions and the solution of the trend conductivity field, which can be calculated once. This idea
is outlined in algorithm~\ref{alg-shadow}.\\

\begin{algorithm}
\DontPrintSemicolon
\tcp{Initialization}
$s = 0$\;
\tcp{Looping over different realizations of $K$}
\For{$i\leftarrow 1$ \KwTo $N$}{
    $K = $ generateField()\;
    \tcp{Generate a large array of uniform random variables to use for the current realization and
    the shadow process}
    $U = $uniformRandArray()\;
    $p_i = $ solveMC($x_0$, $K$, $M=1$, $U$)\;  
    \tcp{Track a shadow particle in the trend conductivity field}
    $p_{shadow} = $solveMC($x_0$, $K_{trend}$, $M=1$, $U$)\;
    $s += p_i - p_{shadow}$\;
}
\tcp{return MC estimate of the mean solution. Here we assume the solution for the trend field, 
$p_{trend}$, is pre-computed.}
\Return $p_{trend}(x_0) + s/N$\;
\BlankLine
 \caption{Using a shadow process for variance reduction in  mean calculation.}
 \label{alg-shadow}
\end{algorithm}

\begin{figure}
\begin{center}
\includegraphics[width=\linewidth]{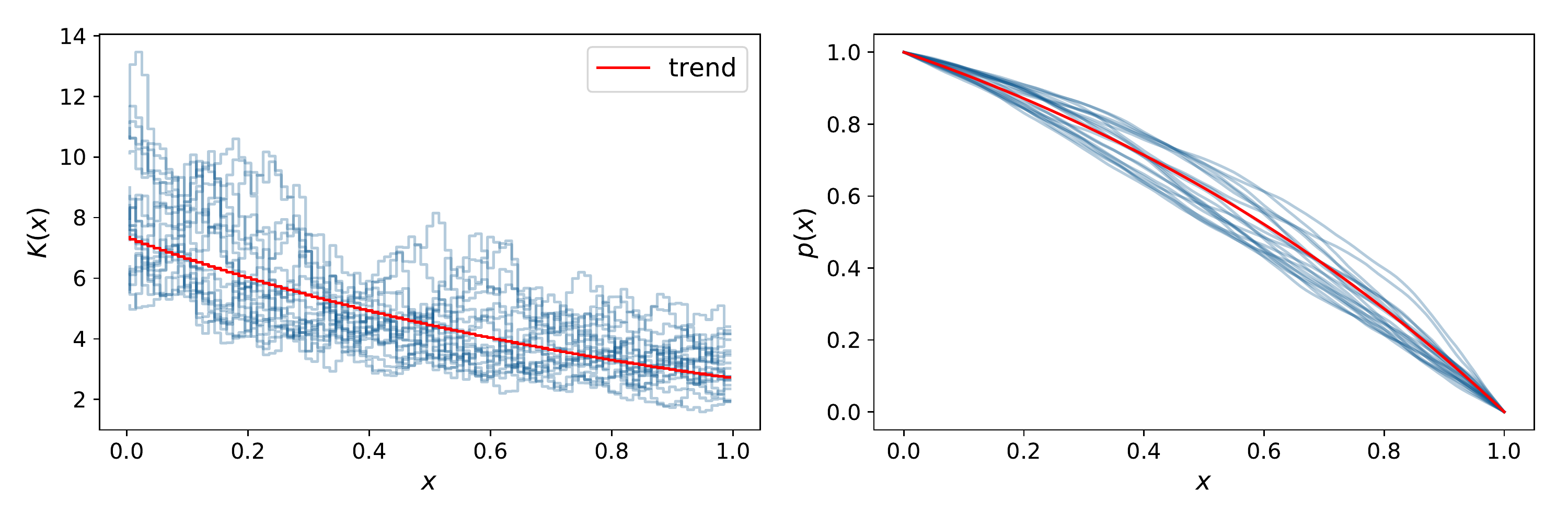}
\caption{Multiple realizations of the conductivity field along with their corresponding pressure
solution for the example in section~\ref{sec-trend}.
\label{trend}} 
\end{center} 
\end{figure}

\begin{figure}
\begin{center}
\includegraphics[width=0.6\linewidth]{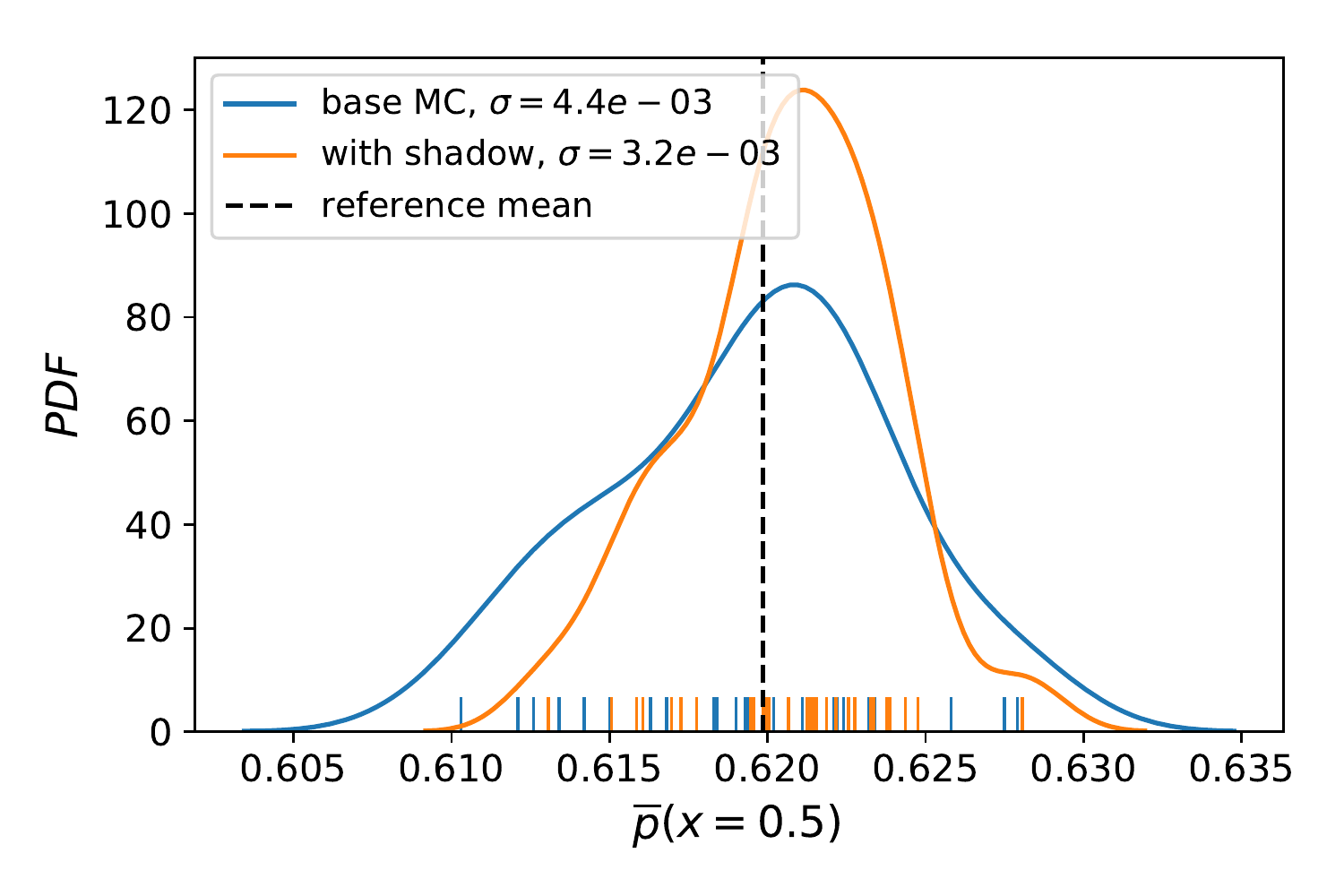}
\caption{The distribution of the mean solution at $x=0.5$ calculated by the base MC
(algorithm~\ref{alg-uq} with $M=1$) and using shadow particles (algorithm~\ref{alg-shadow}) for
thirty different experiments. By using shadow particles, the variance of 
the calculated mean is reduced by $27\%$.\label{var_reduction}}  
\end{center} 
\end{figure}

Here we present an example of variance reduction for calculating the mean in such a setting. 
The mean trend in $\log(K)$ is defined as
\begin{equation}
\log(K)_{trend} = 2-x.
\end{equation}
The log conductivity realizations are generated by adding a Gaussian noise process with mean zero,
standard deviation $\sigma=0.25$ and an exponential correlation structure with $l_Y = 0.25$ to this
trend. A number of conductivity realizations generated using this procedure and their corresponding
pressure solutions are illustrated in Fig.~\ref{trend}.  Figure~\ref{var_reduction} shows the
variance reduction for calculating the mean using shadow particles in the trend conductivity field.
The proposed variance reduction technique works best for cases where the solution to the trend
conductivity field is highly correlated with the solution for different realizations of $K$.  In our
example, this is the case for relatively low variance of the added noise to the log conductivity. 

\section{Conclusions and future work}
\label{sec-conclusions}
In this work, we reviewed the stochastic counterpart of the differential operator in elliptic and
parabolic conservation equations with heterogeneous conductivity fields. Numerical challenges due to
time step restrictions for using this formulation were discussed. A Monte Carlo algorithm is
proposed to solve the elliptic problem for one-dimensional domains with piecewise constant
conductivity. An example was provided to illustrate that this method is capable of accurately
obtaining the solution of a deterministic PDE.  Moreover, the proposed method was used to calculate
accurate estimates of the one point distribution of the solution. It was shown that the
proposed stochastic method can provide a very efficient alternative for estimating the mean 
solution of a random PDE at specific points of interest in the domain. Finally, a variance reduction
scheme was proposed for applying the method for efficient mean calculation.\\

The proposed stochastic simulations can be accelerated using numerical methods designed for the
simulation of stochastic processes. Variance reduction strategies such as 
control variate schemes can be used to decrease the required number of MC trials for a given
precision and will be the subject of future investigations. Moreover, the known statistics of the
conductivity field can be used to accelerate path generation for random walk simulations. In two and
three-dimensional examples that will be the subject of future work, these accelerations can play a
key role. In these higher dimensional domains, extra attention should be given to the simulation of
the random paths close to the boundary (e.g. calculating accurate estimates of the exit locations).
Furthermore, since the random walks only experience the random field locally, generation of complete
realizations of the random field can be avoided. In subsurface flow simulations, this could lead to
significant computational cost savings in generating geostatistical models for uncertainty
quantification.  Finally, an effective implementation strategy to increase the efficiency of the
algorithm is to partition the particle paths and have dedicated cores that simulate paths in each
partition. This implementation strategy will be explored in future investigations.

\section*{Acknowledgments}
Amir H. Delgoshaie is grateful to Daniel M. Tartakovsky and Joseph Bakarji from
the Energy Resources Engineering department at Stanford University for several
helpful discussions.  Funding for this project was provided by the Stanford
University Reservoir Simulation Industrial Affiliates (SUPRI--B) program.

\section*{References}


\end{document}